\pdfoutput=1
\documentclass[11pt]{article}

\usepackage[margin=1in]{geometry}
\usepackage[T1]{fontenc}
\usepackage[utf8]{inputenc}
\usepackage{lmodern}
\usepackage{microtype}
\usepackage{amsmath}
\usepackage{booktabs}
\usepackage{array}
\usepackage{tabularx}
\usepackage{enumitem}
\usepackage{natbib}
\usepackage{url}
\usepackage{xcolor}
\usepackage{hyperref}
\usepackage{tikz}
\usetikzlibrary{positioning, arrows.meta, calc, shapes.geometric, fit, backgrounds}
\usepackage{graphicx}
\usepackage{listings}
\usepackage{caption}
\usepackage{float}

\hypersetup{
  colorlinks=true,
  linkcolor=blue!50!black,
  citecolor=blue!50!black,
  urlcolor=blue!50!black,
  pdftitle={Meta-Engineering Harnesses for AI-Native Software Production},
  pdfauthor={Satadru Sengupta, Tamunokorite Briggs, Ivan Myshakivskyi}
}

\definecolor{hModel}{HTML}{1D1D1F}
\definecolor{hRole}{HTML}{5856D6}
\definecolor{hContract}{HTML}{0071E3}
\definecolor{hMemory}{HTML}{34A853}
\definecolor{hExec}{HTML}{FF9500}
\definecolor{hVerify}{HTML}{BF4800}
\definecolor{hCalib}{HTML}{A2640A}
\definecolor{hMuted}{HTML}{6E6E73}
\definecolor{hLine}{HTML}{86868B}

\newcolumntype{Y}{>{\raggedright\arraybackslash}X}
\setlist[itemize]{leftmargin=*, itemsep=2pt, topsep=3pt}
\setlist[enumerate]{leftmargin=*, itemsep=2pt, topsep=3pt}
\lstdefinestyle{pseudojson}{
  basicstyle=\ttfamily\footnotesize,
  breaklines=true,
  columns=fullflexible,
  keepspaces=true,
  frame=single,
  backgroundcolor=\color{black!2},
  showstringspaces=false,
  tabsize=2
}

\title{\textbf{Meta-Engineering Harnesses for AI-Native Software Production}\\[0.35em]
\large A Contract-Driven Adversarial Verification Architecture with Early Deployment Report}
\author{Satadru Sengupta\\Co-founder \& CEO, HireNimbus\\\texttt{satadru@hirenimbus.com}
\and Tamunokorite Briggs\\Lead Engineer, HireNimbus\\\texttt{tamunokorite@hirenimbus.com}
\and Ivan Myshakivskyi\\Lead Engineer, HireNimbus\\\texttt{ivan@hirenimbus.com}}
\date{}

\begin{document}
\maketitle

\begin{abstract}
AI-native software development is often evaluated at the level of individual models, prompts, or generated artifacts. This framing is insufficient for production environments where software must be continuously produced, verified, deployed, maintained, and adapted across many operational contexts and long time horizons.

We present a meta-engineering harness: a software-production architecture that transforms operational and product feature requirements into explicit contracts, routes work through role-specialized AI agents, performs independent and adversarial verification, and continuously improves itself through structured failure classification and outer-loop calibration.

The harness is designed for settings in which software delivery is not a one-time project but an ongoing operating function. In our motivating application, CTO-as-a-service for small service firms, the system manages websites, booking flows, payment systems, backoffice workflow automations, and AI-agent interfaces as continuously evolving technical infrastructure rather than one-off deliverables.

We describe the layered architecture, including two-pass contract compilation, persistent markdown memory with specialization records, attention-based and independence-based verification, a four-way failure arbiter, and outer-loop calibration. We report results from an early production deployment spanning 17 features over several weeks, including a detailed in-app payments case study that revealed contract incompleteness and verification-boundary issues. These observations directly drove targeted improvements to the harness.

The contribution is an implemented, measurable, and extensible verification architecture for making AI-native service-as-a-software production reliable, auditable, and improvable over time.
\end{abstract}

\section{Introduction}

AI coding agents have made it possible to produce useful software artifacts from natural-language instructions. Recent benchmarks such as SWE-bench \citep{jimenez2023swebench} and SWE-agent \citep{yang2024sweagent} have demonstrated that coding agents can solve real-world GitHub issues autonomously. In the simplest workflow, a human describes a task, an agent writes code, and the human evaluates whether the result appears to work. This workflow is fast, but it is unreliable for production systems. Output quality depends on the prompt, model version, hidden assumptions, available context, and the human's ability to detect subtle failures.

The central problem is not whether an AI model can generate a website, booking workflow, payment integration, or automation once. The central problem is whether a system can continuously produce, verify, deploy, maintain, and upgrade such infrastructure across many operationally distinct but structurally similar businesses.

This paper studies that problem through the lens of a meta-engineering harness: a software-production system around AI agents that governs how requirements become contracts, how contracts become implementations, how implementations are tested, how failures are classified, and how future runs are improved continuously.

The motivating deployment context is CTO-as-a-service for local service firms. Many service businesses need modern software infrastructure but cannot hire a CTO, product manager, engineering team, QA team, or DevOps team. A one-time website or workflow build is insufficient because the infrastructure must evolve as customer expectations, business rules, payment flows, search behavior, AI-agent interfaces, integrations, and security protocols change.

This creates a different technical objective than conventional AI-assisted coding. The objective is not simply to generate artifacts faster. The objective is to build a software-production system that can repeatedly generate, verify, operate, and upgrade business-specific infrastructure.

We make four contributions:
\begin{itemize}
  \item \textbf{A harness-level abstraction for AI-native software production.} We define a meta-engineering harness as a layered architecture around AI agents, contracts, context, verification, deployment, and calibration.
  \item \textbf{A contract-driven adversarial verification architecture.} We describe how operational requirements are compiled into explicit contracts that guide independent implementation, adversarial testing, review, QA, and failure classification.
  \item \textbf{A two-regime verification model.} We distinguish (a) independence-based verification, where separate agents implement and test from the same contract, from (b) attention-based verification, where role-separated reviewers inspect product, architecture, security, QA, and deployment risks.
  \item \textbf{An early deployment case study and evaluation framework.} We report a payments implementation where behavioral correctness was insufficient because missing business logic lived outside the contract, and we use that failure to implement contract refinement, review gates, and harness-level metrics.
\end{itemize}

The paper is intentionally modest in its empirical claims. The case study is diagnostic rather than definitive. It does not prove that the harness generalizes across all software domains. It shows why harness-level verification is necessary, what failure classes emerge in practice, and how those failures can be converted into process improvements.

\section{Problem Setting: From AI Services Arbitrage to Continuous Software Operation}

A common current pattern in AI-native development is technology arbitrage: operators who understand modern coding agents and AI site builders can produce websites, automations, or workflows for customers who do not. This can create short-term value, but it does not solve the harder problem of continuous technical maintenance, upkeep, and upgrade.

Tool knowledge diffuses. Generated outputs commoditize. A one-off build decays as dependencies change, integrations break, security expectations evolve, product requirements shift, and customer workflows become more complex.

For small service firms, the technical surface is broad: web presence, online booking, quoting, scheduling, payments, customer communication, review capture, search and AI-agent discoverability, CRM automation, analytics, integrations, and support workflows.

These businesses typically lack CTOs, product managers, engineers, QA teams, DevOps teams, data teams, and AI automation specialists. This mismatch creates the need for a continuously operated technical layer.

We define CTO-as-a-service as the continuous design, deployment, maintenance, and upgrading of business-specific technical infrastructure on behalf of firms that lack internal technical leadership.

This paper does not treat CTO-as-a-service as a consulting model. It treats it as a software-production problem. The question is whether the production and ongoing upkeep of customer-specific technical infrastructure can be transformed from bespoke labor into a repeatable, verifiable, and continuously improving system.

\section{Meta-Engineering Harness Architecture}

A meta-engineering harness is a software-production architecture that governs how AI agents receive context, transform intent into code, verify outputs, deploy changes, and update future workflows.

The word harness is deliberate. A harness is not a model, prompt, or agent. It is the surrounding system that makes agent outputs more reliable.

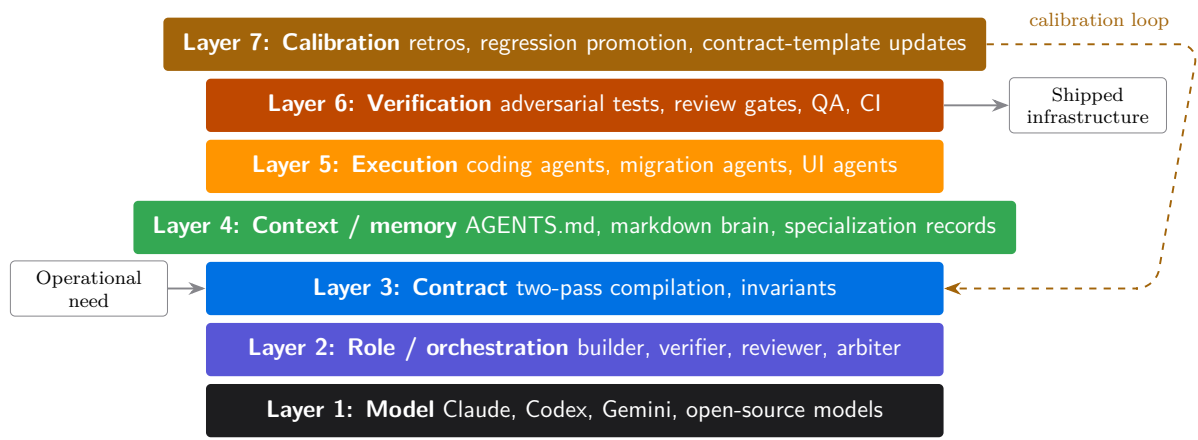
\begin{figure}[H]
\centering
\resizebox{0.96\textwidth}{!}{\begin{tikzpicture}[
    font=\sffamily\small,
    layer/.style={rectangle, rounded corners=2pt, minimum width=10.8cm, minimum height=0.78cm, align=left, text=white, inner xsep=8pt},
    side/.style={rectangle, rounded corners=2pt, minimum width=2.3cm, minimum height=0.7cm, draw=hLine, fill=white, align=center, font=\scriptsize, inner sep=4pt},
    >={Stealth[length=2.5mm]},
    node distance=0.10cm
]
\node[layer, fill=hCalib] (L7) at (0,0) {\textbf{Layer 7: Calibration} \hfill retros, regression promotion, contract-template updates};
\node[layer, fill=hVerify, below=of L7] (L6) {\textbf{Layer 6: Verification} \hfill adversarial tests, review gates, QA, CI};
\node[layer, fill=hExec, below=of L6] (L5) {\textbf{Layer 5: Execution} \hfill coding agents, migration agents, UI agents};
\node[layer, fill=hMemory, below=of L5] (L4) {\textbf{Layer 4: Context / memory} \hfill AGENTS.md, markdown brain, specialization records};
\node[layer, fill=hContract, below=of L4] (L3) {\textbf{Layer 3: Contract} \hfill two-pass compilation, invariants};
\node[layer, fill=hRole, below=of L3] (L2) {\textbf{Layer 2: Role / orchestration} \hfill builder, verifier, reviewer, arbiter};
\node[layer, fill=hModel, below=of L2] (L1) {\textbf{Layer 1: Model} \hfill Claude, Codex, Gemini, open-source models};

\node[side, left=0.55cm of L3] (input) {Operational\\need};
\draw[->, hLine, thick] (input.east) -- (L3.west);

\node[side, right=0.95cm of L6] (output) {Shipped\\infrastructure};
\draw[->, hLine, thick] (L6.east) -- (output.west);

\coordinate (loopTop) at ($(L7.east)+(3.1,0)$);
\coordinate (loopBottom) at ($(L3.east)+(3.1,0)$);
\node[font=\scriptsize, text=hCalib, anchor=south] at ($(L7.east)+(1.65,0.08)$) {calibration loop};
\draw[->, hCalib, thick, dashed, rounded corners=8pt]
    (L7.east) -- (loopTop) -- (loopBottom) -- ++(-0.55,0) -- (L3.east);
\end{tikzpicture}}
\caption{Seven-layer meta-engineering harness. Operational needs enter at the contract layer, where they are compiled into explicit specifications. Execution and verification produce deployable infrastructure, while the calibration layer feeds failure classifications and pattern observations back into contract templates, specialization records, and review gates.}
\label{fig:harness-architecture}
\end{figure}

\begin{table}[H]
\centering
\caption{Layered architecture of the meta-engineering harness.}
\label{tab:layers}
\begin{tabularx}{\textwidth}{p{0.24\textwidth}YY}
\toprule
\textbf{Layer} & \textbf{Function} & \textbf{Example} \\
\midrule
Model layer & Selects or swaps models & Claude, Codex, Gemini, open-source models \\
Role/orchestration layer & Assigns cognitive roles and execution order & compiler, backend agent, frontend agent, tester, reviewer, arbiter \\
Contract layer & Converts intent into executable specification & GitHub issue, JSON contract, markdown spec \\
Context/memory layer & Maintains persistent project state & AGENTS.md, markdown brain, specialization records \\
Execution layer & Produces implementation artifacts & coding agents, migration agents, UI agents \\
Verification layer & Checks behavior and system fit & adversarial tests, review gates, QA, CI \\
Calibration layer & Learns from failures & retros, regression promotion, contract-template updates \\
\bottomrule
\end{tabularx}
\end{table}

Attention-based, or role-based, orchestration in multi-agent systems has been explored in frameworks such as MetaGPT \citep{hong2023metagpt} and AutoGen \citep{wu2023autogen}. This layering separates concepts that are often conflated:
\begin{itemize}
  \item A prompt is a single instruction.
  \item Context is the information assembled around the instruction.
  \item An agent is a model operating within a constrained role.
  \item A harness is the system that controls prompts, context, roles, tools, verification, and feedback.
  \item A software factory is the larger production system: harness, accumulated contracts, memory, specialization registry, test suites, deployment infrastructure, and calibration history.
\end{itemize}

The harness is the software-production architecture that makes the factory run.

\subsection{Implementation Status}

The harness has been implemented as a production workflow rather than only described as a conceptual framework. The current implementation uses GitHub issues, compiled contracts, role-specific AI workers, adversarial test generation, CI runners, human arbitration, markdown memory, and retrospective calibration. Over the initial deployment window, the system was applied to 17 features across backend services, mobile flows, provider websites, payments, scheduling, notifications, product pages, and search-tool integration.

The current implementation is not fully autonomous by design. Human operators still approve high-stakes contract changes, classify failures the arbiter cannot reliably resolve, and review permanent memory updates. The design goal is not to remove human judgment, but to relocate it from repetitive implementation toward contract design, exception handling, and harness governance.

\section{Contracts, Context, and Specialized Agents}

\subsection{Contracts}

Building on spec-driven development \citep{piskala2026spec} and test-driven agent definition \citep{rehan2026tdad}, the harness uses a contract as its central data structure. It converts a raw request into a specification precise enough for one agent to build against and another agent to verify.

A contract should encode module name, user roles, API or UI surface, desired behavior, inputs and outputs, state transitions, invariants, business rules, error taxonomy, authentication and authorization rules, data dependencies, out-of-scope clauses, QA targets, regression risks, and acceptance criteria.

Without a contract, the implementation agent is forced to infer missing requirements. With a contract, the builder, verifier, reviewer, and QA process share a source of truth.

\subsection{Two-Pass Contract Compilation}

The harness uses a two-pass compilation process. Pass 1 is a completeness pass: the raw issue is converted into a structured draft, making implicit assumptions explicit through types, state transitions, edge cases, trust boundaries, and error conditions. Pass 2 is a scope and ambiguity pass: the draft is reduced and clarified, unsupported requirements are removed, and ambiguous clauses are rewritten so downstream agents do not treat multiple interpretations as equally valid.

This second pass was introduced after observing that first-pass contracts can over-specify. Over-specification is dangerous because downstream agents treat unsupported requirements as mandatory.

\subsection{Persistent Context and Memory}

A major failure mode in AI-native development is context evaporation, a consequence of LLMs being stateless between calls. Decisions made in one session do not reliably carry into the next.

The harness uses repository-level markdown memory to preserve relevant project context. The memory structure has two sections:
\begin{itemize}
  \item Permanent section: human-approved institutional knowledge that automated processes cannot modify directly.
  \item Rolling section: recent pattern observations subject to compression, promotion, or deletion.
\end{itemize}

The goal is not perfect memory. The goal is controlled compression: preserving decisions, constraints, and recurring failure patterns likely to affect future software production.

\subsection{Specialization Records}

The harness maintains a specialization registry for recurring task domains, such as payments, booking, authentication, search, reviews, or mobile frontend flows. A specialization record can inject domain-specific requirements into contract compilation and downstream review. For example, a payments specialization may require idempotency keys, explicit state transitions, and trust-boundary checks.

Specializations are applied only when confidence crosses a threshold. Below that threshold, the pipeline proceeds without the specialization to avoid corrupting the contract with incorrect domain assumptions.

\subsection{Specialized Agents}

Each agent has a bounded role, an approach consistent with role-based agent architectures in prior work \citep{hong2023metagpt,wu2023autogen}. The contract compiler does not implement the feature. The implementation agent does not write the adversarial tests. The review agent did not write the implementation.

This reduces role contamination and makes failures easier to classify. At larger scale, specialization also reduces prompt length, context contamination, output variance, unnecessary file access, and broad refactors.

\section{Verification: Independence-Based and Attention-Based}

The harness uses two complementary verification regimes.

\subsection{Independence-Based Verification}

In independence-based verification, one agent implements from the compiled contract while another agent writes tests from the same contract without seeing the implementation, building on test-driven agent definition \citep{rehan2026tdad} and adversarial multi-agent testing approaches \citep{thukkaram2025adversarial}.

The goal is to reduce implementation-specific confirmation bias. A verifier that did not write the code is less likely to adopt the implementation's assumptions.

The harness enforces independence structurally through separate agents, separate job payloads, separate execution queues, no shared conversation history, and verifier access to the contract rather than implementation reasoning, following the pattern of adversarial Red-Blue agent architectures \citep{thukkaram2025adversarial} and Bayesian adversarial multi-agent frameworks \citep{zeng2026bayesian}.

This is not formal independence. Current LLM APIs do not provide proof that two agents have no correlated biases. If both foundation models are trained on similar data or both receive an incomplete contract, they may share blind spots. Structural independence reduces correlated implementation blindness; it does not eliminate correlated model failure.

\subsection{Attention-Based Verification}

In attention-based verification, the same or a similar model is directed into different reviewer roles: product reviewer, architecture reviewer, security reviewer, backend reviewer, frontend reviewer, QA tester, and shipping reviewer. This extends role-based agent specialization from prior work \citep{hong2023metagpt} into a multi-pass review regime.

This is not true independence. Its purpose is different: it changes the inspected surface.

Independence-based verification reduces implementation blindness. Attention-based verification reduces single-pass attentional blindness.

\begin{table}[H]
\centering
\caption{Complementary verification regimes in the harness.}
\label{tab:verification}
\begin{tabularx}{\textwidth}{p{0.22\textwidth}YYY}
\toprule
\textbf{Verification type} & \textbf{Mechanism} & \textbf{Catches} & \textbf{Limitation} \\
\midrule
Independence-based & Separate builder and verifier & contract violations, behavioral bugs & misses contract gaps and shared blind spots \\
Attention-based & Role-separated review & product, architecture, UX, structural issues & not formally independent \\
Human outer loop & Human judgment and failure classification & ambiguous intent, business judgment, process failures & costly and must become exception-based at scale \\
\bottomrule
\end{tabularx}
\end{table}

\subsection{Four-Way Failure Arbiter}

When tests fail, the arbiter classifies the failure into one of four categories.

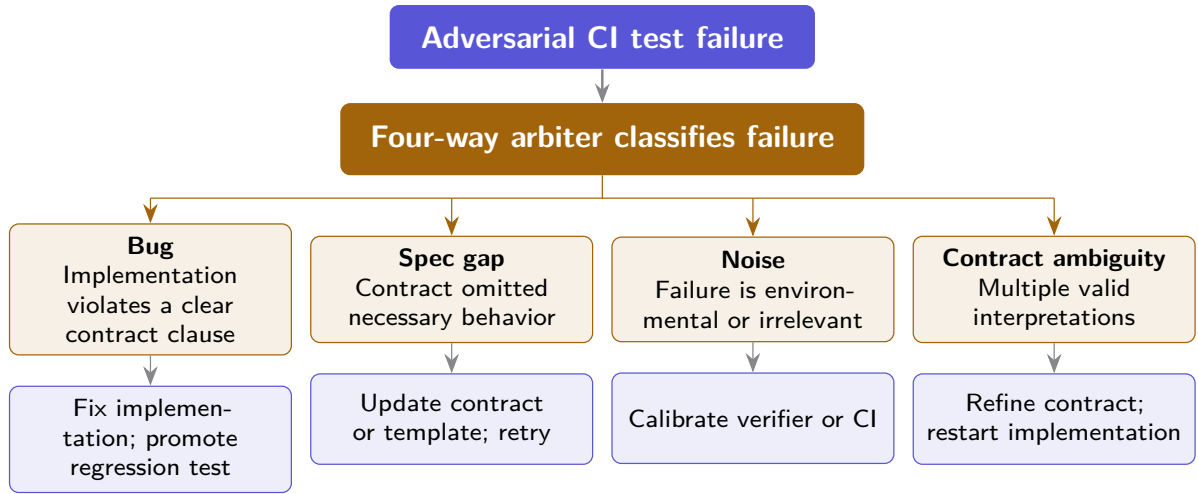
\begin{figure}[H]
\centering
\resizebox{0.96\textwidth}{!}{\begin{tikzpicture}[
    font=\sffamily\scriptsize,
    >={Stealth[length=2.2mm]},
    input/.style={rectangle, rounded corners=3pt, fill=hRole, text=white, minimum width=4.8cm, minimum height=0.72cm, align=center, font=\sffamily\small\bfseries},
    arb/.style={rectangle, rounded corners=3pt, draw=hCalib, fill=hCalib, text=white, minimum width=5.9cm, minimum height=0.8cm, align=center, font=\sffamily\small\bfseries},
    class/.style={rectangle, rounded corners=3pt, draw=hCalib, fill=hCalib!10, text width=2.9cm, minimum height=1.18cm, align=center, inner sep=4pt},
    action/.style={rectangle, rounded corners=3pt, draw=hRole, fill=hRole!10, text width=2.9cm, minimum height=1.02cm, align=center, inner sep=4pt}
]
\node[input] (fail) at (0,0) {Adversarial CI test failure};
\node[arb, below=0.38cm of fail] (arbiter) {Four-way arbiter classifies failure};
\draw[->, hLine, thick] (fail) -- (arbiter);

\node[class] (bug)   at (-5.1,-2.85) {\textbf{Bug}\\Implementation violates a clear contract clause};
\node[class] (spec)  at (-1.7,-2.85) {\textbf{Spec gap}\\Contract omitted necessary behavior};
\node[class] (noise) at ( 1.7,-2.85) {\textbf{Noise}\\Failure is environmental or irrelevant};
\node[class] (amb)   at ( 5.1,-2.85) {\textbf{Contract ambiguity}\\Multiple valid interpretations};

\coordinate (rail) at (0,-1.82);
\draw[hCalib] (arbiter.south) -- (rail);
\draw[hCalib] (-5.1,-1.82) -- (5.1,-1.82);
\draw[->, hCalib] (-5.1,-1.82) -- (bug.north);
\draw[->, hCalib] (-1.7,-1.82) -- (spec.north);
\draw[->, hCalib] ( 1.7,-1.82) -- (noise.north);
\draw[->, hCalib] ( 5.1,-1.82) -- (amb.north);

\node[action, below=0.34cm of bug] (abug) {Fix implementation; promote regression test};
\node[action, below=0.34cm of spec] (aspec) {Update contract or template; retry};
\node[action, below=0.34cm of noise] (anoise) {Calibrate verifier or CI};
\node[action, below=0.34cm of amb] (aamb) {Refine contract; restart implementation};

\draw[->, hLine] (bug) -- (abug);
\draw[->, hLine] (spec) -- (aspec);
\draw[->, hLine] (noise) -- (anoise);
\draw[->, hLine] (amb) -- (aamb);
\end{tikzpicture}}
\caption{Four-way failure arbiter. Every adversarial CI failure is classified before corrective action is taken. Wrong classification produces wasted cycles; in particular, contract ambiguity should trigger contract refinement rather than implementation retry.}
\label{fig:arbiter}
\end{figure}

\begin{table}[H]
\centering
\caption{Failure classes and corrective actions.}
\label{tab:arbiter}
\begin{tabularx}{\textwidth}{p{0.22\textwidth}YY}
\toprule
\textbf{Failure class} & \textbf{Meaning} & \textbf{Corrective action} \\
\midrule
Bug & Implementation violates a clear contract & fix implementation; promote regression test \\
Spec gap & Contract omitted necessary behavior & update contract/template; retry \\
Noise & Failure is environmental or irrelevant & calibrate verifier or CI \\
Contract ambiguity & Contract allows multiple valid interpretations & refine contract; restart implementation \\
\bottomrule
\end{tabularx}
\end{table}

The contract ambiguity category is especially important. If a contract admits multiple valid behaviors, retrying the implementation is wasteful. The correct response is contract refinement.

\section{Pipeline and Outer-Loop Calibration in Three Phases}

The harness pipeline has three phases.

\subsection{Pre-Pipeline}
\begin{enumerate}
  \item Operational need arises.
  \item Raw issue is drafted.
  \item Contract compiler produces structured contract.
  \item Product review checks whether this is the right thing to build.
  \item Engineering review checks state transitions, data dependencies, architecture, and failure modes.
  \item Contract is finalized.
\end{enumerate}

\subsection{Pipeline}
\begin{enumerate}[resume]
  \item Implementation agent receives compiled contract.
  \item Test agent receives compiled contract.
  \item Implementation agent writes code.
  \item Test agent writes adversarial tests.
  \item CI runs tests against implementation.
  \item Failures are routed to the arbiter.
  \item Bug/spec/noise/ambiguity classification determines next action.
\end{enumerate}

\subsection{Post-Pipeline}
\begin{enumerate}[resume]
  \item Structural review checks architecture, repo patterns, trust boundaries, performance, and maintainability.
  \item QA checks staging, browser behavior, API behavior, or mobile flow behavior.
  \item Shipping workflow deploys or opens a merge request.
  \item Retro agent reviews failure history and proposes harness updates.
  \item Human approves permanent memory and specialization changes.
\end{enumerate}

The pipeline does not stop at deployment. Each failure is an observation about the harness. A recurring bug may require a new regression test. A recurring spec gap may require a contract template update. A recurring review failure may require a new checklist item. A recurring ambiguity may require a new compiler rule.

This is the calibration loop. The harness improves by converting failures into reusable process changes.

\section{Case Study: In-App Payments Backend}

\subsection{Task}

The case study involved implementing in-app payments for a homeowner mobile application using Stripe PaymentIntents and Stripe Connect across a backend Lambda service and a React Native frontend. The backend required endpoints for creating payment intents, verifying payment completion, managing invoice state transitions, and preserving trust boundaries around client-reported payment status.

\subsection{Contract}

The contract specified endpoints, authentication rules, Stripe behavior, state transitions, invariants, error taxonomy, side effects, and terminal payment states. Appendix~\ref{app:contract} includes a redacted excerpt of the compiled contract. One important clause specified that the payable amount should be derived from \texttt{quote\_data.total} and should not be recomputed from other fields.

\subsection{Pipeline Result}

The backend implementation reached passing CI in two cycles. From the standpoint of the contract and adversarial tests, the implementation was behaviorally correct. However, two failures emerged later.

First, a final invoice payment did not correctly deduct deposits that had been paid outside Stripe and marked manually. The implementation calculated deposit totals using only the Stripe payments table. Second, discount calculations were not correctly applied to the final Stripe amount.

Both failures traced to contract incompleteness. The contract correctly constrained basic payment flows with no discounts or deposits, but it did not encode the full business logic required for final invoice calculation or more complex regular payment scenarios. The harness built what the contract specified, but the contract did not fully capture the intended behavior.

\subsection{Frontend Result}

The frontend was implemented in one attempt but encountered React Native dependency and environment issues that the automated pipeline could not classify. Human intervention was required. Manual review also caught missing permission configurations and navigation-state handling.

\subsection{Lessons}

The case study illustrates two failure modes. Contract incompleteness means the implementation can satisfy the contract while failing the true business requirement. Verification boundary means an adversarial test suite conditioned on the contract cannot catch behavior outside the contract.

These failures motivated two harness additions: contract compilation and refinement, and structured review gates. The key lesson is that passing behavioral tests is necessary but insufficient. Production software must also fit the surrounding business and technical system.

\section{Evaluation Metrics and Early Operational Evidence}

Evaluation must occur at the harness level, not only the model level.

\subsection{Metrics}

\begin{table}[H]
\centering
\caption{Harness-level evaluation metrics.}
\label{tab:metrics}
\begin{tabularx}{\textwidth}{p{0.20\textwidth}Y}
\toprule
\textbf{Metric class} & \textbf{Examples} \\
\midrule
Productivity & features completed; cycle time per feature; autonomous commits; human interventions; cost per shipped change; average implementation cycles \\
Verification & adversarial test suites generated; pass/fail rate; bugs caught before merge; spec-gap rate; verifier-noise rate; ambiguity detection rate; regression tests promoted \\
Quality & post-merge bug rate; rollback rate; staging QA failure rate; production incident rate; structural review findings; trust-boundary findings \\
Calibration & recurring failure classes; contract sections associated with gaps; agents associated with failures; specialization updates; memory promotions; failures converted into reusable improvements \\
\bottomrule
\end{tabularx}
\end{table}

\subsection{Diagnostic Metrics}

\begin{enumerate}
  \item \textbf{Contract violation detection rate:} fraction of implementation violations caught before merge.
  \item \textbf{Review gate precision:} fraction of review-gate failures that correspond to real issues rather than false positives.
  \item \textbf{Average implementation cycles:} mean number of attempts per feature.
  \item \textbf{Ambiguity detection rate:} frequency with which ambiguous contracts are correctly routed to contract refinement rather than implementation retry.
\end{enumerate}

\subsection{Current Evidence and Caveats}

Over 3--4 weeks of deployment across a service-business context, the harness implemented 17 features: a force-update modal, in-app payments, a scheduling module, a product landing page, an MCP search tool integration, a Slack notification workflow, 6 provider websites, and several bug fixes. 18 adversarial test suites were generated across these features, plus 15 additional suites produced during iterative calibration of the scheduling module. 5 bugs or implementation gaps were caught before merge, including a missing field in a Slack notification integration and a codebase-standards violation.

\begin{table}[H]
\centering
\caption{Early operational evidence from the initial deployment window.}
\label{tab:evidence}
\begin{tabularx}{\textwidth}{p{0.40\textwidth}Y}
\toprule
\textbf{Evidence item} & \textbf{Observed value} \\
\midrule
Deployment window & 3--4 weeks \\
Features implemented & 17 \\
Adversarial test suites generated & 18 \\
Additional calibration suites & 15 \\
Bugs / implementation gaps caught pre-merge & 5 \\
Backend payment cycles to passing CI & 2 \\
Known post-CI business-logic misses & 2 \\
Human intervention categories & CI environment, cache keys, large-file context, website refactors \\
\bottomrule
\end{tabularx}
\end{table}

The payments case study provides the most diagnostic evidence. The backend passed CI in two cycles, but the adversarial test suite missed two business-logic failures, a deposit deduction and a discount calculation, because neither was encoded in the contract. The harness built what the contract specified; the contract did not fully capture the intended behavior.

Three further failure patterns emerged from the broader deployment. One bug fix attempt could not be completed because the relevant code files were large and poorly documented, beyond what the contract and context layers could reliably cover. One implementation required manual cache-key edits to finalize a fix the harness had partially resolved. In 2--3 provider website implementations, generated code required manual refactoring before passing deployment checks.

These results reflect an early harness in active use, not a controlled experiment. They identify where the system worked, where it fell short, and what the failure patterns suggest about where investment in contract quality and verification coverage will have the most impact.

\section{Limitations and Threats to Validity}

\subsection{Contract Incompleteness}

The harness is only as good as the contract. If a critical requirement is missing, the builder may not implement it and the verifier may not test it. Contract completeness is the highest-leverage unsolved problem in the system.

\subsection{Shared Model Blind Spots}

Adversarial independence is structural, not formal. Separate agents may still share biases, training-distribution assumptions, or systematic misreadings.

\subsection{Verification Coverage}

Tests sample behavior. They do not prove correctness. Contract-conditioned tests cannot catch failures outside the contract.

\subsection{Human Bottlenecks}

Some decisions require human judgment, including product intent, trust boundaries, ambiguous tradeoffs, and failure classification. At scale, human interventions must become exception-based.

\subsection{Context Drift}

Persistent markdown memory can become stale, bloated, or contradictory. Compression reduces the risk but does not eliminate it.

\subsection{Cost and Latency}

Multi-agent workflows are more expensive and slower than direct model calls. A complete run may take several minutes, and parallelization is not always possible.

\subsection{Security}

Agents with tool access create new attack surfaces. A malformed issue, compromised specialization record, or unsafe tool invocation could influence downstream behavior.

\subsection{Evaluation Difficulty}

It is difficult to isolate the effect of the harness from model improvements, human expertise, task selection, team familiarity with the system, or founder/operator involvement. The deployment evidence comes from a single organization and a proprietary codebase, with no randomized baseline. The results should therefore be interpreted as early operational evidence rather than externally validated performance claims.

The harness does not eliminate human judgment. It changes where human judgment is applied: from repetitive implementation toward contract design, exception handling, calibration, and governance.

\section{Related Work}

This paper sits at the intersection of several areas:
\begin{itemize}
  \item \textbf{AI-native programming:} coding agents, IDE copilots, benchmark-driven code generation, and autonomous software-engineering agents \citep{chen2021codex,jimenez2023swebench,yang2024sweagent}.
  \item \textbf{Specification-driven development:} approaches that use explicit specifications, contracts, invariants, and tests to guide implementation \citep{piskala2026spec,rehan2026tdad}.
  \item \textbf{Software verification and testing:} test-driven development, property-based testing, fuzzing, CI/CD, static analysis, and code review automation \citep{fucci2017tdd,thukkaram2025adversarial,zeng2026bayesian}.
  \item \textbf{Multi-agent systems:} role-based agents, debate, reflection, tool-using agents, and multi-agent orchestration \citep{hong2023metagpt,wu2023autogen}.
  \item \textbf{Human-in-the-loop systems:} failure classification, postmortems, retrospectives, approval gates, and process calibration.
  \item \textbf{Organizational learning and software process improvement:} mechanisms by which repeated failures become persistent process changes through failure classification, retrospective review, and outer-loop calibration.
\end{itemize}

We do not claim that contracts, CI, code generation, QA, role-based agents, or retrospectives are individually novel. The contribution is their integration into a contract-driven verification architecture for continuously operated AI-native software production.

\section{Artifact Availability and Reproducibility}

The production implementation described in this paper is proprietary and cannot be fully released in its current form. The paper therefore provides a redacted contract excerpt, pipeline description, failure taxonomy, evaluation metrics, and case-study analysis sufficient to reproduce the architecture at the workflow level. A future reference implementation could release a minimal harness demonstrating contract compilation, independent implementation/test workers, failure classification, and markdown memory without exposing production code or customer data.

Because the current artifact is not public, the empirical claims are limited to early operational evidence. The paper should be read as an architecture and experience report rather than a benchmark paper.

\section{Discussion}

The case study suggests a broader distinction between artifact generation and infrastructure operation. AI-native coding can produce artifacts quickly. But maintaining business-specific technical systems over time requires more than generation. It requires persistent memory, explicit contracts, independent verification, review gates, deployment checks, and feedback loops.

In this model, humans do not disappear from the software-production process. Their role shifts. Humans become contract designers, product-intent judges, exception handlers, calibration supervisors, and governance owners.

For small service firms, the implication is practical: technical capability that previously required internal engineering teams or expensive agencies may become deliverable as a continuously operated service via software. For software teams, the implication is organizational: engineering work shifts from writing every line of code to designing, supervising, and improving the production system that writes, tests, deploys, and updates code.

The durable asset is not a generated website, booking flow, or payment integration. It is the accumulated production system: contracts, specialization records, failure taxonomies, regression suites, customer-specific context, workflow templates, QA targets, deployment infrastructure, and calibration history. Each shipped feature, failed test, human correction, and production incident can improve the system's future performance.

\section{Conclusion}

AI-native coding has shown that foundation models can produce useful software artifacts quickly. For one-off tasks, that may be sufficient. For continuously operated software infrastructure, it is not.

This paper introduced the meta-engineering harness as a contract-driven verification architecture for AI-native software production. The harness combines compiled contracts, persistent context, role-specialized agents, adversarial verification, role-separated review, QA gates, failure classification, and outer-loop calibration.

The main claim is that reliability should be evaluated at the level of the harness, not the individual model call. A model can generate an artifact. A harness can make production measurable, auditable, and improvable over time.

For CTO-as-a-service, this architecture marks the difference between short-lived technology arbitrage and a scalable operating model built on an adaptive production system. The goal is not to build a website or workflow once. The goal is to build a production system that can repeatedly build, verify, operate, and improve technical infrastructure across many businesses.

\appendix
\section{Redacted Contract Excerpt, Payments Case Study}
\label{app:contract}

This appendix presents a redacted pseudo-JSON excerpt of the compiled contract. It is not intended to be parsed as valid JSON; production identifiers, environment names, and internal implementation details have been removed.

\begin{lstlisting}[style=pseudojson]
{
  "module": "NGPayments",
  "version": "1.0.0",
  "api": {
    "base_path": "/ng/payments",
    "auth": "Supabase Bearer token; JWT identity authoritative",
    "endpoints": [
      {
        "method": "POST",
        "path": "/ng/payments/intent/:invoiceId",
        "returns": ["client_secret", "payment_intent_id", "payment_type", "amount_cents"],
        "side_effects": [
          "store payment_intent_id on invoice",
          "set invoice.payment_status = processing"
        ],
        "errors": ["403", "404", "409", "422", "502"]
      },
      {
        "method": "POST",
        "path": "/ng/payments/confirm/:invoiceId",
        "stripe_verification": "retrieve PaymentIntent and assert status == succeeded before DB write",
        "side_effects_on_success": [
          "set invoice payment status",
          "set service request status",
          "set paid_at",
          "clear payment_intent_id"
        ],
        "side_effects_on_failure": [
          "reset invoice.payment_status to unpaid",
          "clear payment_intent_id"
        ]
      },
      {
        "method": "GET",
        "path": "/ng/payments/status/:invoiceId",
        "returns": ["payment_status", "payment_type", "amount_cents", "currency", "paid_at"]
      }
    ]
  },
  "invariants": [
    "PaymentIntent uses transfer_data.destination",
    "no platform fee",
    "server verifies Stripe success before DB write",
    "paid is terminal",
    "processing blocks duplicate intents",
    "amount derived from quote_data.total"
  ],
  "known_gap_identified_after_deployment": [
    "final invoice calculation omitted offline deposits",
    "discount calculation not encoded in original contract"
  ]
}
\end{lstlisting}

\end{document}